\documentclass[preprint2]{aastex6}

\usepackage{hyperref,ifthen,psfig,graphicx}
\usepackage{epstopdf,placeins}
\usepackage{soul,color,amsmath}
\usepackage{ifthen,psfig,graphicx}
\usepackage{epstopdf,aas_macros}
\usepackage{soul,color,amsmath}
\def\mathnew{\mathsurround=0pt}
\def\simov#1#2{\lower 2.5pt\vbox{\baselineskip0pt \lineskip-.5pt
\ialign{$\mathnew#1\hfil##\hfil$\crcr#2\crcr\sim\crcr}}}
\def\simless{\mathrel{\mathpalette\simov <}}
\def\simgreat{\mathrel{\mathpalette\simov >}}
\newcommand{\MeV}{Me\kern-0.11em V}
\newcommand{\keV}{ke\kern-0.11em V}

\newcommand{\es}{erg~s$^{-1}$}
\newcommand{\kms}{\ensuremath{\rm km\;s}^{-1}}
\newcommand{\cgsbri}{\ensuremath{\rm{erg\;cm}^{-2}\;\rm{s}^{-1}\rm{\AA}^{-1}\rm{arcsec}^{-2}}}
\newcommand{\omegam}{$\Omega_{m,0}$\/}
\newcommand{\omegal}{$\Omega_{\Lambda,0}$\/}

\makeatletter
\newcommand\iont[2]{{#1$\;${\small\expandafter\@slowromancap\romannumeral #2@\relax}}}
\makeatother
\makeatletter
\newcommand{\raisemath}[1]{\mathpalette{\raisem@th{#1}}}
\newcommand{\raisem@th}[3]{\raisebox{#1}{$#2#3$}}
\makeatother

\slugcomment{Accepted by ApJ; Published 2016 September 21}

\shorttitle{Mapping Excitation Modes in NGC 3393}
\shortauthors{Maksym et al.}

\begin{document}


\title{Mapping Seyfert and LINER Excitation Modes in the Inner kpc of NGC 3393}


\author{W. Peter Maksym, Giuseppina Fabbiano, Martin Elvis, \\ Margarita Karovska, Alessandro Paggi, John Raymond}
\affil{Harvard-Smithsonian Center for Astrophysics, \\ 60 Garden St., Cambridge, MA 02138, USA}
\email{walter.maksym@cfa.harvard.edu; @StellarBones}

\author{Junfeng Wang}
\affil{Department of Astronomy, Physics Building, Xiamen University, \\ Xiamen, Fujian, 361005, China}

\and

\author{Thaisa Storchi-Bergmann}
\affil{Departamento de Astronomia, Universidade Federal do Rio Grande do Sul, \\ IF, CP 15051, 91501-970 Porto Alegre, RS, Brazil}



\begin{abstract}

We have mapped the extended narrow line region (ENLR) of NGC 3393 on scales of
$r\lesssim4\arcsec$ ($\sim1\,$kpc) from the nucleus using emission line images of
H$\alpha\,\lambda6563$,
[\ion{O}{3}]$\lambda5007$, and {[\ion{S}{2}]$\lambda\lambda6717,6731$} taken with $HST$ as
part of the {\it CHandra survey of Extended Emission line Regions in nearby
Seyfert galaxies} ({\it CHEERS}).
By mapping these lines onto a spatially resolved Baldwin-Phillips-Terlevich (BPT)
diagram, we investigate the impact of feedback from a Compton-thick AGN on its circumnuclear
ISM.  We find the expected Seyfert-like emission within the ionization {bicone} ($\lesssim3\arcsec$; 770 pc).  
We also find a new, figure 8 shaped LINER cocoon enveloping the bicone and defining a sharp ($\simless100\;$pc) transition between higher and lower ionization zones.  These data illustrate the morphological dependence of ionization states of the ENLR relative to bicone and host gas geometries.

\end{abstract}




\section{Introduction}

\begin{table*}[ht]

\caption{Hubble Observation Properties}
\label{table:hst-obstable}
\centering
\begin{tabular}{cccccc}
\tableline
\tableline

Dataset	&	Obs Date	&	Exposure (s)	 & Instrument & Filter & Note	\\
%		&							&\multicolumn{3}{c}{($\times10^{-14}$\ecmss)}	\\
\tableline
IBIG06050	&	2011 Nov 11	& 2040	&	WFC3/UVIS	&  F814W & I-band	\\
\tableline
IBLY01011	&	2011 May 16	& 566	&	WFC3/UVIS	&  FQ508N & [\ion{O}{3}]	\\
IBLY01021	&	2011 May 17	& 466	&	WFC3/UVIS	&  F665N & H$\alpha$+[\ion{N}{2}]	\\
IBLY01GWQ	&	2011 May 17	& 208	&	WFC3/UVIS	&  F547M & line continuum	\\
IBLY01GXQ	&	2011 May 17	& 208	&	WFC3/UVIS	&  F621M & line continuum	\\
IBLY01GYQ	&	2011 May 17	& 314	&	WFC3/UVIS	&  F673N & [\ion{S}{2}]	\\

\tableline

\end{tabular}
%% Any table notes must follow the \end{tabular} command.\\
\tablenotetext{}{IBIG exposures are from program 12185 (PI: Greene). \\ IBLY exposures are from CHEERS, program 12365 (PI: Wang). }
\end{table*}

Active Galactic Nuclei (AGN) seem to play a critical role in the evolution of their host galaxies.  As the supermassive black hole (SMBH) accretes material, photoionizing radiation and kinetic outflows can regulate AGN accretion and galactic star formation via positive and negative feedback, which in turn affects the rate of SMBH growth \citep{SR98,Fabian12,HB14}.  The ability to quantify the relative impact of different modes of feedback is of interest for characterizing this process, but at large distances from the observer the inability to resolve the extended narrow line region (ENLR) limits its usefulness as a tool for characterizing AGN behavior.   Detailed studies of the ENLR in nearby galaxies are therefore necessary to help understand these processes in detail \citep[see e.g.][]{SB10,Wang11b,Wang11c,Paggi12,Barbosa14}.

Low Ionization Emission Line Regions \citep[LINERs;][]{Heckman80} are a puzzling feature of AGN and starbursts as they may be excited by several processes \citep{Ho08}.  Without a means to disentangle these mechanisms, these common \citep{NB16} objects cannot be used to cleanly investigate either AGN or starbursts.

\citet*{BPT81} pioneered the use of narrow line ratio diagnostic diagrams in the classification of AGN, using ratios between a variety of lines such as [\ion{N}{2}]$\,\lambda6584$/H$\alpha\,\lambda6563$ and [\ion{O}{3}]$\,\lambda5007$/H$\beta\,\lambda4861$ to describe the observed differences between various extragalactic phenomena in terms of activity that is primarily Seyfert-driven, LINER-like, or star-formation-driven %(i.e. by \ion{H}{2} regions; 
(see \citealt{Kewley01, Kewley06}).  This figure is now called the ``BPT diagram".  BPT diagrams have become important tools for classifying large samples of galaxies.  

It is also possible to make spatially resolved BPT diagrams, such that different parts of a galaxy are mapped in terms of their BPT classification \citep[e.g.][for a recent example using Integral Field Units; IFUs]{Cresci15, Davies16, Ho14}.  

We exploit this method in this paper to see if LINER emission has some special morphology.  We use the higher, space-based, angular resolution of the {\it Hubble Space Telescope} ({\it HST}) with narrow band filters.  Low-redshift galaxies present a particular opportunity due to the advantages  of angular scale.

Here we investigate the bright \citep[$m_B=13.1$;][]{deV91} nearby face-on spiral galaxy NGC 3393 as part of  {\it CHandra survey of Extended Emission line Regions in nearby Seyfert galaxies} ({\it CHEERS}; PI: J. Wang).  NGC 3393 is a low-luminosity \citep[$L_X\rm{[0.5-10 keV]}\sim10^{41}$\;\es;][]{Levenson06} type 2 Seyfert galaxy at $z=0.0125$ \citep[$D=53\;$Mpc;][]{Theureau98}.  NGC 3393 has a water maser disk \citep{Kondratko08}, X-rays show it is Compton-thick \citep{Maiolino98}, and various studies imply $L_{bol}\sim10^{44}$\;\es\ \citep[within an order of magnitude; see][]{Kondratko08,Winter12,Baumgartner13,Koss15}.  The host galaxy has long been known to exhibit evidence for complicated jet-NLR interactions, with a significant ENLR covering much of the galaxy, dominated by S-shaped emission line ``arms" which wrap around compact sub-kpc radio jet lobes \citep{Cooke00}.  

\cite{Cooke00} investigated the ENLR morphology using pre-COSTAR [\ion{O}{3}]$\lambda5007$ and H$\alpha$+[\ion{N}{2}] $\lambda\lambda6563,6548,6584$ images, and produced a sparsely sampled resolved BPT diagram of the entire galaxy using single-slit spectroscopy.  Here we use more recent (post-COSTAR) {\it HST} narrow-band images obtained via {\it CHEERS} to investigate the ENLR excitation on $\sim10\,$kpc scales using a resolved BPT diagram and image map covering the region within $\sim8\arcsec$ of the nucleus.  

This is the first paper in a series, to be followed by an in-depth study of the comparative optical, X-ray and radio morphology of the NGC 3393 ENLR, as well as a study of its resolved narrow-line X-ray morphology.  In this series, we describe the physical mechanisms impacting the excitation and evolution of the NGC 3393 ENLR in detail that complements and extends beyond the recent work of \cite{Koss15}. 

Throughout this paper, we adopt concordant cosmological parameters\footnote{Distances are calculated according to http://www.astro.ucla.edu/$\sim$wright/CosmoCalc.html} of
$H_0=70\ $km$^{-1}$ sec$^{-1}$ Mpc$^{-1}$, \omegam=0.3 and \omegal=0.7. All coordinates are J2000.  For distances, we use $D=53\;$Mpc, such that NGC 3393 presents an angular scale $256\;\rm{pc\;arcsec}^{-1}$.

\section{Observations and Data}

\subsection{HST Observations}

NGC 3393 has been observed extensively by {\it HST} over its mission, including WFPC1, WFPC2, FOS, STIS, ACS and NICMOS.  For this investigation we mainly use observations taken for  {\it CHEERS} (program 12365).    {\it CHEERS} observations were taken using the Wide Field Camera 3 (WFC3) instrument in a sub-array configuration of the UVIS channel on 16 May 2011 and 17 May 2011 using the FQ508N, F665N, F547M, F621M, and F673N filters.  We also used observations taken with the F814W filter on 2011 November 11 (Program 12185, PI: Greene).  These short observations lasted 
566s, 466s, 208s, 208s, 314s and 2040s respectively, and are listed in Table  \ref{table:hst-obstable} along with relevant bandpasses.  

At the redshift of NGC 3393, our narrow filter observations cover $\sim42\;$\AA\ (observer frame) for each of three bands, with FQ508N, F665N and F673N covering [\ion{O}{3}]$\lambda5007$, H$\alpha\,\lambda6563$ with [\ion{N}{2}]$\lambda\lambda$6548,6584, and the  [\ion{S}{2}]$\lambda\lambda$6716,6731 doublet respectively.  We measure the the [\ion{O}{3}]  continuum using the F547M band, and the H$\alpha$+[\ion{N}{2}] and [\ion{S}{2}] continuum from F621M.  We expect all of the line and continuum bands to be free of contamination from other significant emission lines.

\subsection{Data Reduction}

\begin{figure*}
\noindent
\includegraphics[width=0.32\textwidth]{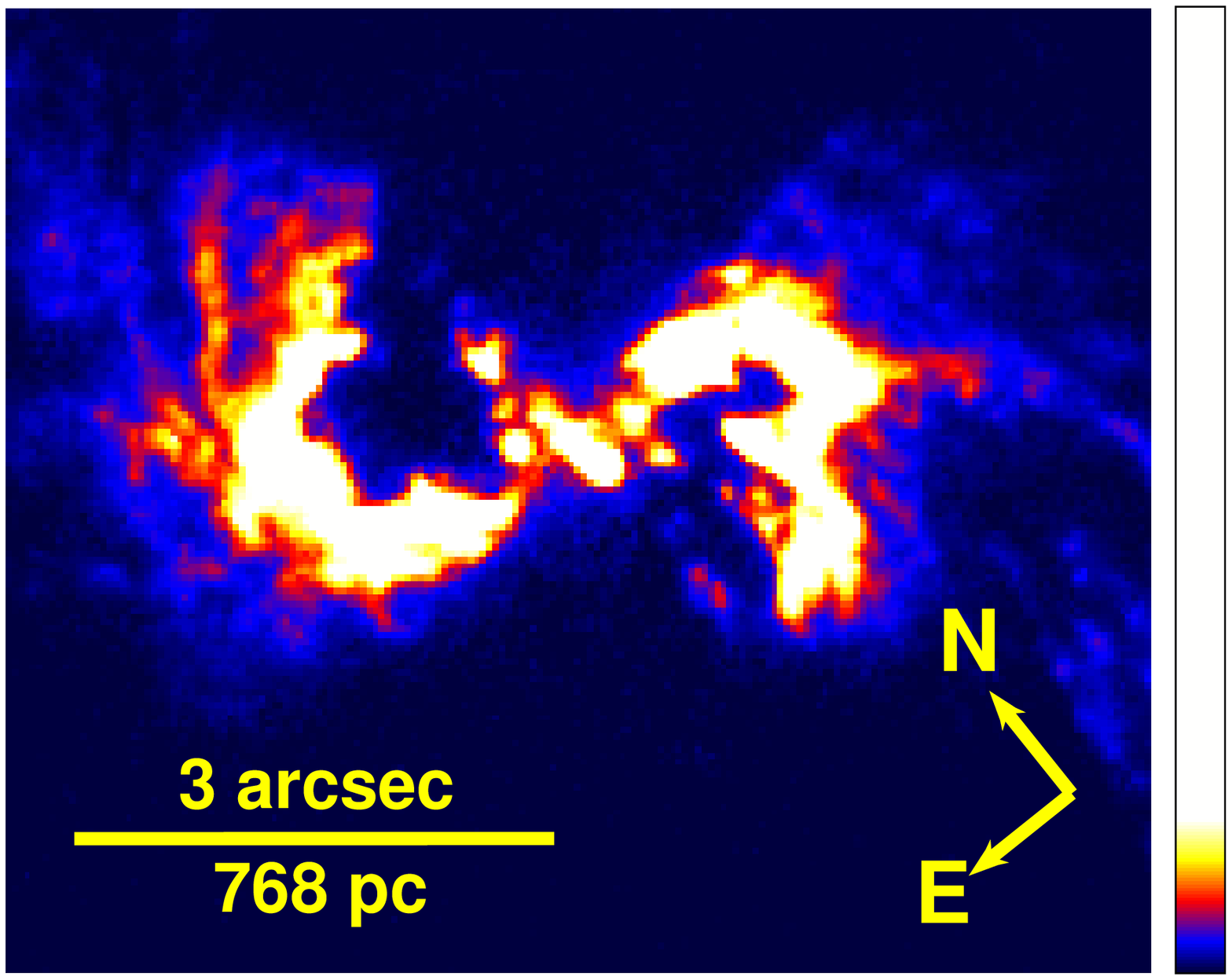}\hspace{0.02\textwidth}%
\includegraphics[width=0.32\textwidth]{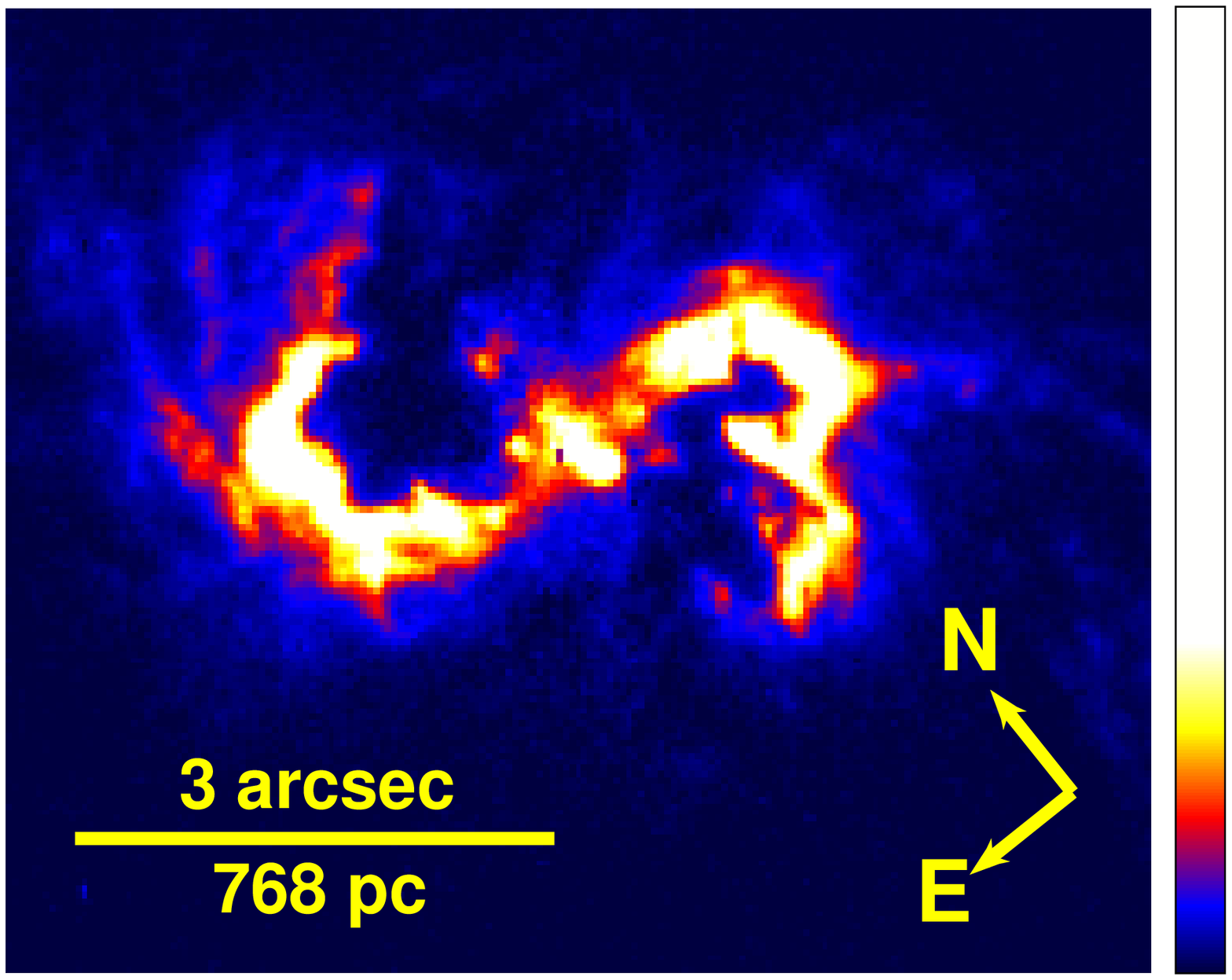}\hspace{0.02\textwidth}%
\includegraphics[width=0.32\textwidth]{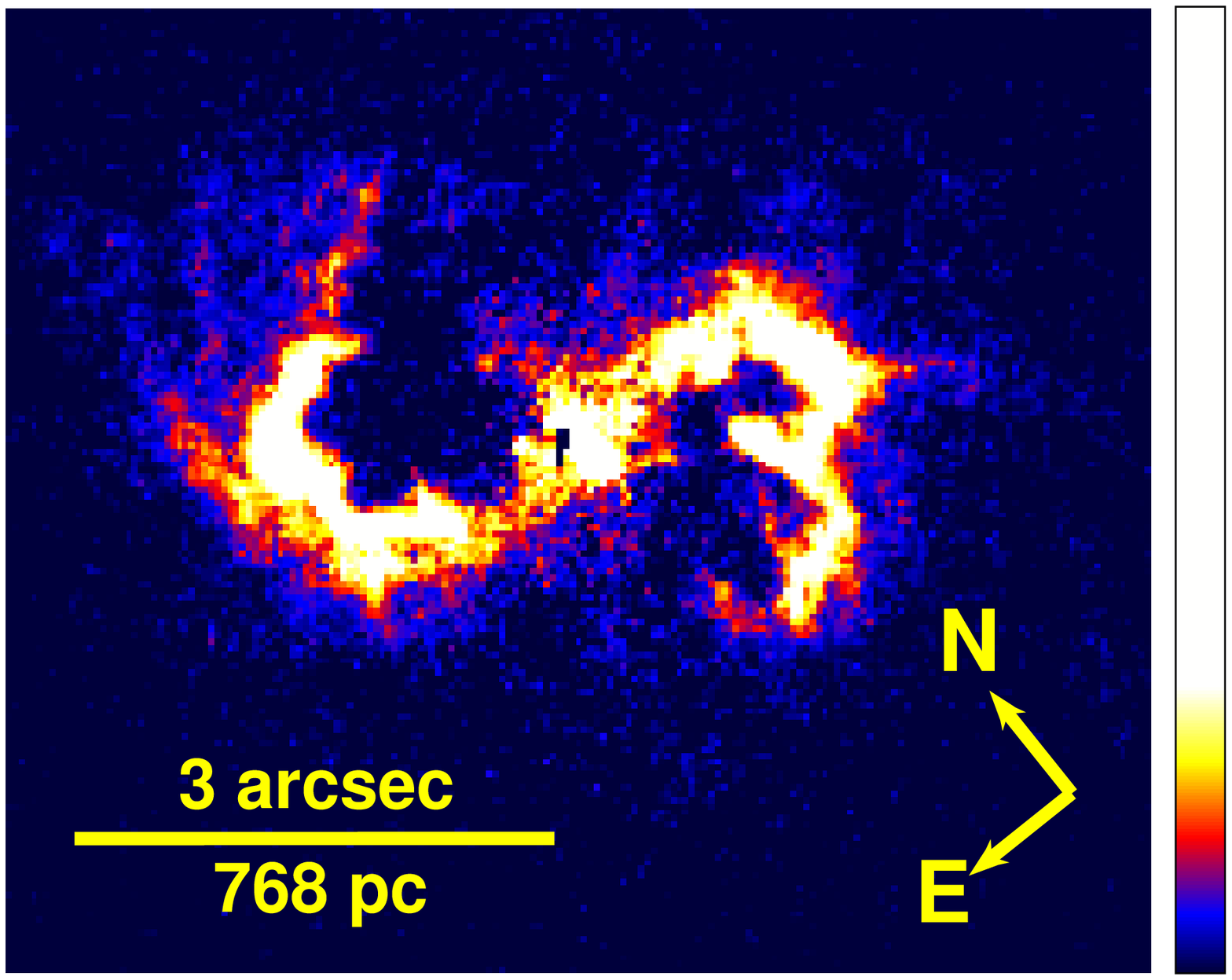}\par
\caption{Continuum-subtracted surface brightness images of  [\ion{O}{3}] (left), H$\alpha$ (middle) and [\ion{S}{2}] (right).  Images are blanked at 3$\sigma$ significance.  In H$\alpha$ and [\ion{S}{2}], there is a flaw near the nucleus a few pixels in size due to poor cosmic ray subtraction in the continuum band.
} 
\label{fig:line-sb}
\end{figure*}

We used the standard {\it HST} data processing package DrizzlePac \citep{astrodrizzle} to process {\it HST} images with Pyraf.  This software was installed with version 1.5.1 of Ureka\footnote{http://ssb.stsci.edu/ureka/}.  Cosmic rays and bright sources can leave faint residual charge trails in the direction of readout.  Software is available for for WFC3 CTE correction\footnote{http://www.stsci.edu/hst/wfc3/ins\_performance/CTE/}, but is not capable of handling the subarray configuration of most of our images, and is therefore not used here.  The most and longest observations are taken using the F814W filter, with correspondingly better cosmic ray removal and catalog source significance.  We therefore use the F814W images for astrometric reference.  In each band, we reject cosmic rays using {\tt AstroDrizzle}.  We match sub-exposure astrometry using DrizzlePac tools {\tt TweakReg} and {\tt TweakBack}, and use {\tt AstroDrizzle} again to produce a combined band image, varying the source catalog and cosmic ray criteria as necessary.  We match the astrometry in different bands to the final F814W images using  {\tt TweakReg} and {\tt TweakBack}.  We then match the pixel sampling to the native pixel scale ($\sim0.04\arcsec$ width) and the orientation of the F814W images using {\tt AstroDrizzle}.

Observations with filters F547M, F621M, and F673N have only single exposures, which eliminates the possibility of cosmic ray removal by direct comparison within a single filter band.  We therefore use the final F814W image as a reference for cosmic ray removal in F547M and F621M.  We use the F665N image as a reference for cosmic ray removal in the F673N band.  Differences in surface brightness gradients when comparing images in different bands can create false cosmic ray detections for sufficiently aggressive cosmic ray removal criteria.  We therefore use criteria which leave a number of cosmic rays, and remove them manually with the Pyraf tool {\tt imedit}.  We also remove several cosmic rays from the FQ508N image using {\tt imedit}.

Background determination is complicated by the large size of NGC 3393 relative to the fields of view of the WFC3 observations.  This can cause background over-subtraction in the AstroDrizzle median sigma-clipping routine.  The darkest continuum regions, for example, exhibit some flux when viewed at 3.6$\mu$m with {\it Spitzer} (Program 100098; PI: Stern).  We therefore subtract a background component which matches the spatial gradients (in regions dominated by old stellar populations) seen in {\it Spitzer}.  The resulting background is consistent with median values of the darkest regions to $\simless10\%$ for all images, and is less than the uncertainty in the observed continuum.  Where the line flux $\geq3\sigma$, this background is no more than a few percent of line flux.

\subsection{Narrow Line Mapping}
\label{sec-lines}

We produced emission line maps of [\ion{O}{3}]$\lambda5007$, H$\alpha\,\lambda6563$ and [\ion{S}{2}] $\lambda\lambda$6716,6731 by rescaling and subtracting the {\it HST} continuum bands from the corresponding narrow line filter observations.  The CIAO tool {\tt dmimgcalc}\footnote{http://cxc.harvard.edu/ciao/ahelp/dmimgcalc.html} is capable of generic numerical manipulations of FITS image files, and was used for this purpose.  We rescaled these images according to the WFC3 absolute flux calibration keyword PHOTFLAM, and by pixel size in $\rm{arcsec}^2$, to have images in units of \cgsbri.  We adjust the scaling of the continuum prior to subtraction, such that regions with both high-S/N continuum and negligible expected line emission (i.e. outside the ionization cones) produce median continuum-subtracted flux values consistent with zero.

To map [\ion{O}{3}]$\lambda5007$, we simply rescaled the F547M image to match FQ508N regions outside the ionization cones and inner $\sim2\,$kpc, then subtracted the result from FQ508N, assuming excess flux is due solely to  [\ion{O}{3}]$\lambda5007$.  H$\alpha\,\lambda6563$ and [\ion{S}{2}] $\lambda\lambda$6716,6731 require more complicated modeling due to variation of filter transmission and contamination by [\ion{N}{2}]$\lambda\lambda6548,6584$.  We subtract the F621M continuum similarly from F665N and F673N.  We then model the respective contributions of different lines' emission to the total count rate in filters using {\tt pysynphot} from Ureka to simulate the expected count rates in F665N and F673N.  According to \cite{Cooke00}, [\ion{N}{2}] varies little with radius on spatial scales of $\simless1\arcsec$.  We therefore assume Gaussian emission profiles with FWHM $336\;\rm{km\;s}^{-1}$ and line strengths according to Table 7 in  \cite{Cooke00}, taken from ground-based spectroscopy.

In order to correct for line filling from H$\alpha$ absorption in the host galaxy starlight, we use methods based on \cite{Keel83}.  We extrapolate the relative continuum strength at 5500\,\AA\ and 6700\,\AA\ from F547M and F621M and use {\tt pysynphot} and {\tt dmimgcalc} to calculate fluxes.  We correct H$\alpha$ for continuum absorption from the expected equivalent width based on the continuum slope as per \cite{Keel83}.  We use the CIAO tool {\tt dmimgthresh} and {\tt Pyraf} to limit the range of H$\alpha$ absorption equivalent width to between 0\,\AA\ and 2.6\,\AA, which is the largest value considered by \cite{Keel83}.  The brightest features in the nuclear ENLR do exceed 2.6\,\AA, suggesting estimates of the stellar continuum in the arms for F547M and F621M are strongly affected by local gas, possibly due to  [\ion{O}{1}]$\lambda\lambda6000,6300$ emission falling within F621M.  Using the \cite{Keel83} relationship between color and absorption, we find $H\alpha$ filling to be a minor effect, $\simless\rm{few}\;\%$.  The final line continuum-subtracted line maps are presented in Figure \ref{fig:line-sb}.  These images are blanked at 3$\sigma$ \citep[as per][]{Cresci15} to emphasize regions which are sufficiently bright for line ratio mapping at maximum resolution without additional processing.

\section{BPT Mapping}
\label{sec:BPTmap}

\begin{figure}
\centering
\includegraphics[width=0.47\textwidth]{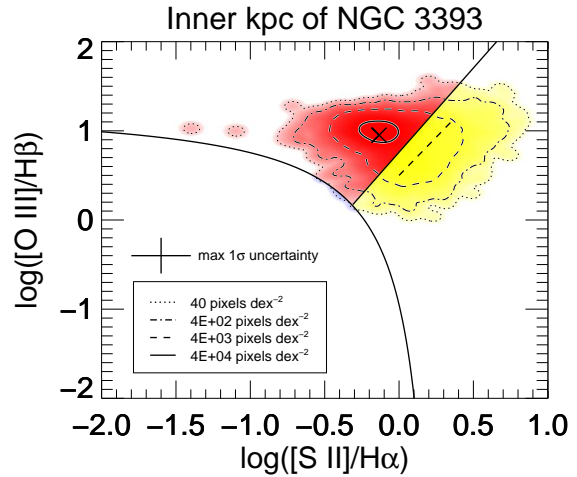} \\
\caption{BPT Diagram of the inner kpc radius region of NGC 3393, covering a $\sim7\arcsec\times7\arcsec$ NE-SW region centered on the nucleus and ionization cones.  As in Figure \ref{fig:BPT-img}, {\bf Red} corresponds to Seyfert-like activity (top-left), {\bf Yellow} corresponds to LINER-like activity (right), and {\bf Blue} pixels have line ratios typical of \ion{H}{2} regions (bottom-left). Solid black lines mark the boundaries of these regions according to Kewley et al (2006).  Contours and intensity of color indicate phase space density, in terms of number of WFC3 pixels per square dex in the BPT diagram where $\rm{S/N}\geq3\sigma$.  Most \ion{H}{2}-like pixels are excluded by this criterion.  An `X' marks the integrated fluxes where $r<2\arcsec$ from the nucleus.  The cross indicates the $1\sigma$ uncertainty of any pixel which is only detected at $3\sigma$ in all lines.  A dashed line indicates the elongated parameter space overdensity, mentioned in the text but not visible in this color stretch.}
\label{fig:BPT-chart}
\end{figure}

\begin{figure}
\centering
\includegraphics[width=0.47\textwidth]{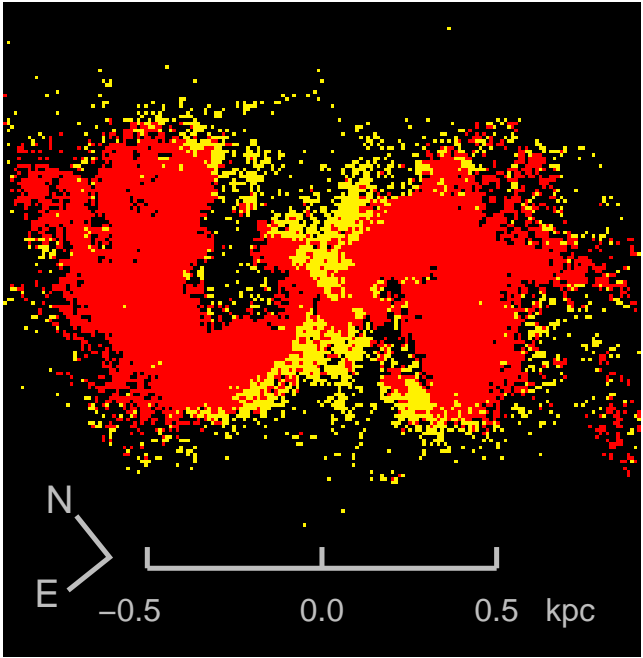} \\
\caption{Resolved BPT image of the inner kpc radius region of NGC 3393, covering a $\sim7\arcsec\times7\arcsec$  NE-SW region ($7\arcsec\sim1.8\;$kpc) centered on the nucleus and ionization cones.  Each WFC3 pixel has been colored according to its position on the BPT diagram in Figure \ref{fig:BPT-chart} relative to its [\ion{S}{2}]/H$\alpha$ and [\ion{O}{3}]/H$\beta$ ratios.  {\bf Red} corresponds to Seyfert-like activity and {\bf Yellow} corresponds to LINER-like activity.  {\bf Black} pixels have at least one line where  $F_{line}\leq3\sigma$ in this region. {\bf Blue} pixels represent line ratios typical of \ion{H}{2} regions, and are entirely excluded here due to weak line significance. 
}
\label{fig:BPT-img}
\end{figure}

We use the formalism of \cite{Kewley06} to produce resolved BPT maps of the inner $\sim$kpc radius region ($\sim4\arcsec$) of NGC 3393.  In order to do this, we divide pairs of emission line maps from \S\ref{sec-lines} and Fig. \ref{fig:line-sb} to create ratio maps of [\ion{S}{2}]/H$\alpha$ and [\ion{O}{3}]/H$\beta$ at the native WFC3 pixel size.  A value of $\rm{H}\alpha\sim3\times\rm{H}\beta$ has long been shown to be typical of AGN NLRs \citep{HS83,GF84}, and is consistent with the \cite{Cooke00} value for NGC 3393.  We therefore use H$\alpha$ as a proxy for H$\beta$ according to this ratio.  

In Figure \ref{fig:BPT-chart}, we plot a resolved BPT diagram of log([\ion{O}{3}]/H$\beta$) vs. log([\ion{S}{2}]/H$\alpha$), such that each $0.04\arcsec\times0.04\arcsec$ WFC3 pixel ($10\,\rm{pc}\times10\,\rm{pc}$) represents one data point on the BPT diagram.  The extraction region covers a $\sim7\arcsec\times7\arcsec$ region aligned NE to SW which includes the nucleus (center) and the ionization cones.  We use $3\sigma$ blanking as a requirement for emission line significance \citep[as in][]{Cresci15}.

Figure \ref{fig:BPT-chart} shows all the $3\sigma$-blanked image pixels in log([\ion{O}{3}]/H$\beta$) vs. log([\ion{S}{2}]/H$\alpha$) space.  Contours of pixel density are shown.  The cross marks the peak, which lies in the Seyfert region, in agreement with line ratios from the integrated nuclear region at $r<2\arcsec$.  
A secondary elongated structure is in the LINER region (not visible in this color stretch), oriented along the LINER/Seyfert divide, such that [\ion{O}{3}]/H$\beta$ increases along with [\ion{S}{2}]/H$\alpha$.

%\placefigure{fig:BPT-chart}

We can examine the location of different emission line excitation classes by displaying an image of the circumnuclear region such that each WFC3 pixel is color-coded according to its position on the BPT diagram.  Figure \ref{fig:BPT-img} shows the BPT image of NGC 3393. Red indicates Seyfert-type activity; yellow is LINER-like, and blue is typical of star-forming regions.  Pixels without at least $\sim3\sigma$ detection are excluded (black).  

%\placefigure{fig:BPT-img}
%\FloatBarrier

We immediately see several important features of the morphology. 

1) Within $r\sim3\arcsec$ ($\sim770\,$pc) of the nucleus, the ionization cones are dominated by an S-shaped structure filled with Seyfert-type emission (red).  
 
2) Narrow line emission within $r\sim3\arcsec$ but outside the cones is predominantly unclassified due to weak line significance (particularly in [\ion{S}{2}]).  Our $3\sigma$ blanking criterion (black in Fig. \ref{fig:BPT-img}) selects against low-emission regions, which may also select against \ion{H}{2}-like regions at the lines' limiting fluxes (Fig. \ref{fig:BPT-img}; blue).
   
3) The interface between these two spatial regions is an open figure-8 of LINER-type emission (yellow in Fig. \ref{fig:BPT-img}).  This interface is thin, $\lesssim100\,$pc ($\lesssim0.4\arcsec$) in projection.  This LINER-like `cocoon' is a newly-identified structure in this galaxy.

\section{Discussion}\label{discuss}

Narrow line imaging of [\ion{S}{2}] and reasonable assumptions about H$\alpha$/H$\beta$ and H$\alpha$/[\ion{N}{2}] (described in \S\ref{sec:BPTmap}) have allowed us to study the excitation of the ENLR on scales of $\sim10\;$pc by placing it on the BPT diagram.  Figure \ref{fig:BPT-img} shows a very clean morphology for the inner $\sim770\;$pc ($\sim3\arcsec$).  

1) Since the ionization cone interiors are almost entirely comprised of Seyfert-like emission, we may simply suppose their origin is predominantly by photoionization.  However, if the S-shape is formed by fast shocks with photoionized precursors \citep[as considered by][]{Cooke00}, shock velocities of $\simgreat500\;\kms$ can produce photoionization-like BPT positions \citep{Allen08}.  

2) The cross-cone region emission becomes too faint to classify $\sim300\;$pc from the nucleus.  This limits our ability to measure the full extent of the LINER cocoon, which may be much thicker than $\sim100\,$pc.  Deeper observations are necessary to show if the cocoon is bounded by a lower-ionization \ion{H}{2}-like region \citep[as in][]{Cresci15}.

3) The LINER-like `cocoon' may be due to shock production \citep{Heckman80,Allen08}, or could consist of regions with large local gas densities relative to the AGN ionizing flux (hence their low ionization), or regions receiving diluted radiation due to filtering of the nuclear radiation \citep{Kraemer08,HB14} due to shielding by the the edge of the torus or by a hollow biconical AGN wind, each of which may create a zone of lower ionization.

\cite{Kraemer08} distinguish two ionization states in the base of the NGC 4151 wind out to $\sim70\;$pc using [\ion{O}{3}]/[\ion{O}{2}].  Using only two lines, however, they can only distinguish high and low ionization states.  Our analysis is consistent with that picture, such that the addition of a third line gives a cleaner classification.  We resolve larger-scale structure in finer detail, however (10-pc pixels at all scales here, vs. wedges of 11-pc to 44-pc angular extent at $r\sim30-130\,$pc in NGC 4151), allowing us to see the thin LINER cocoon.  Even with an additional diagnostic line for BPT analysis, the $20\degr$ azimuthal bins used by \cite{Kraemer08} would be too large to resolve this LINER-like region, and each pixel in our map is comparable ($\sim10\;$pc) to one of their radial bin steps.

The LINER-like emission is located in a `cocoon' surrounding the Seyfert emission regions.  Shocks resulting from gas expanding outward from the ionization cone with $v\simgreat200\,\kms$ could explain these regions as the result of the AGN wind pushing into the surrounding ISM.  

Whether or not shocks play a significant role, the origin of the LINER-like cocoon could be analogous to the `Fermi bubbles' in our own Milky Way galaxy, assuming those bubbles were driven by an AGN wind in the recent past \citep{Su10}.

The detailed morphology of the ENLR suggests that different morphologies of obscuration and gas distribution may affect the classification of a galaxy taken as a whole, e.g. at high redshift or in a SDSS $3\arcsec$ fiber \cite{Kewley13}.  Modeling is necessary to fully investigate the effects of plausible geometries (but is beyond the scope of this paper).  

%In the case of NGC 3393, the combination of wide ionization cones ($\simgreat100\degr$) intersecting the spiral disk structure provides bright gas with a high [\ion{O}{3}]/H$\alpha$ ratio out to $r\sim1\,$kpc. The flux may be due to material highly ionized by some combination of shocks at the leading edge of the radio jet and AGN wind, and photoionization by the bicone at $\simless0.5\;$kpc \citep{Cooke00}.

\section{Conclusions}

Using narrow-band {\it HST} images, we have produced a resolved BPT diagram and BPT map of a $\sim7\arcsec\times7\arcsec$ ($\sim1.8\;$kpc square) region covering the nucleus and inner ionization cones of NGC 3393. We have shown that the regions within the bicones are typically consistent with Seyfert galaxies, consistent with previous classification of the galaxy based on the nucleus alone, while the line emission outside these regions has line surface brightness too low to categorize at full resolution.  The surprise is that we find a thin LINER-like `cocoon' bounding the AGN-like region.  Deeper observations (such as with {\it HST} and {\it JWST}) are necessary to measure the cocoon's full extent and answer the question: Is the cocoon as thin as it seems?

Our ability to distinguish these different regions is limited by the availability of different diagnostic lines, and by the resulting degeneracies between the models within the observed parameters.  

The BPT maps are consistent with the LINER cocoon at the edge of the bicone being shielded by the torus or AGN wind, or with shocks.  

Further investigation of these mechanisms is therefore necessary.  In the next paper of this series (Maksym et al., 2016, {\it in preparation}), we will investigate the limits of the broad set of available observational data for NGC 3393, including a more detailed study of the optical morphology in conjunction with high-resolution radio and X-ray data.  That work will provide the basis for a detailed study of {\it Chandra} emission line imaging data.  Given the role that optical narrow line ratios play in understanding the ENLR, we expect X-ray mapping of important diagnostic lines like \ion{Ne}{9}, \ion{O}{7} and \ion{O}{8} to provide a similarly  important, complementary, role in understanding ENLR processes in AGN, as demonstrated by \cite{Wang11b}, \cite{Wang11c} and \cite{Paggi12}.  A deeper understanding of the morphology on sub-kpc scales will better inform methods of X-ray spectral extraction, which is critical given the limitations due to low photon counting rates in the high-resolution regime of X-ray imaging.

%% XXX COMMENT ON KS14



%% Included in this acknowledgments section are examples of the
%% AASTeX hypertext markup commands. Use \url without the optional [HREF]
%% argument when you want to print the url directly in the text. Otherwise,
%% use either \url or \anchor, with the HREF as the first argument and the
%% text to be printed in the second.

\acknowledgments

WPM acknowledges support from {\it Chandra} grants GO4-15107X, GO5-16099X, and GO2-13127X, and {\it Hubble} grant GO-13741.002-A, and thanks Henrique Schmitt for helpful discussions.  J.W. acknowledges support from the NSFC grants 11473021 and 11522323, and the Fundamental Research Funds for the Central Universities under grant 20720160023.  We acknowledge partial support by NASA contract NAS8-03060 (CXC).  STSDAS and PyRAF are products of the Space Telescope Science Institute, which is operated by AURA for NASA.  We thank the anonymous referee for comments which improved the quality of the paper.  ME and GF thank the Aspen Center for Physics, funded by NSF grant \#1066293, for their hospitality while this paper was completed.

\facility{HST,Spitzer}

\end{document}